      [1999/12/01 v1.4c Il Nuovo Cimento]
\documentclass{cimento}

\usepackage{graphicx}
\title{Broadband modeling of gamma-ray burst afterglows}
\author{Alexander~J.~van~der~Horst\from{ins:1}, Ralph~A.M.J.~Wijers\from{ins:1} \atque Evert~Rol\from{ins:2}}
\instlist{\inst{ins:1} Astronomical Institute ``Anton Pannekoek'', University of Amsterdam
  \inst{ins:2} University of Leicester}
\PACSes{\PACSit{98.70.Rz}{gamma-ray sources; gamma-ray bursts}
\PACSit{95.30.Gv}{Radiation mechanisms; polarization}
\PACSit{95.85.Bh}{Radio, microwave ($>1$ mm)}}
\begin{document}

\maketitle

\begin{abstract}
According to the fireball model gamma-ray burst (GRB) afterglows are the result of a shock pushed into the surrounding medium by an extremely relativistic outflow from the GRB. By modeling the broadband spectrum, ranging from X-ray to radio wavelengths, on time scales of hours to years after the GRB, we can determine several burst parameters, e.g. the burst energy, the circumburst medium density (and its gradient) and the energy contained in electrons and the magnetic field. We have developed a code to do broadband modeling and determine the model parameters. We also performed Monte Carlo simulations with synthetic data sets in order to derive an estimate of the accuracy of the best fit parameters. 

GRB970508 was taken as a case study and we present our fitting results. We also present our WSRT radio observations of GRB030329 at centimeter wavelengths and the results from modeling this afterglow. The low frequency WSRT radio observations provide a unique possibility to study the late stages of the fireball evolution and examine the jet structure.
\end{abstract}

\section{Broadband modeling}
The dominant radiation mechanism for GRB afterglows is synchrotron emission. The synchrotron spectrum is determined by the peak flux and three break frequencies, namely the synchrotron self-absorption frequency $\nu_a$, the frequency  $\nu_m$ that corresponds to the minimal energy in the electron energy distribution, and the cooling frequency $\nu_c$ that corresponds to electrons that lose their energy quickly by radiation. The evolution of the spectrum is determined by the hydrodynamical evolution of the blastwave propagating into the circumburst medium. The break frequencies and the peak flux can be described in terms of the energy of the blastwave $E$, the density of the surrounding medium $n$, the fractional energy densities behind the relativistic shock in electrons and in the magnetic field, $\epsilon_e$ and $\epsilon_B$ respectively, the slope $p$ of the electron energy distribution, and the jet opening angle $\theta_{jet}$. 

We developed a code to combine the theoretical spectra and light curves with observational data from GRB afterglows by using chi-square fitting \cite{ref:1}. We can put in a broadband data set and fit it simultaneously in time and frequency, and from that determine the model parameters. We also perform Monte Carlo simulations with synthetic data sets in order to derive an estimate of the accuracy of the best fit parameters. 

We modeled GRB970508 as a case study. In our model we assume that the density of the surrounding medium as a function of the distance has a power law profile, i.e. $n\propto r^{-k}$, and in our code we treat $k$ as an independent parameter. For GRB970508 we find that $k=0.0010\pm 0.0005$, which means that the blast wave expands into a medium with a uniform density. For the other fit parameters we find $E=(2.23\pm 0.32)\cdot 10^{51}\,\mbox{erg}$, which is the beaming corrected energy, $n=0.081\pm 0.040\,\mbox{cm}^{-3}$, $\epsilon_e=0.24\pm 0.04$, $\epsilon_B=0.29\pm 0.12$, $p=2.15\pm 0.04$ and $\theta_{jet}=0.80\pm 0.06\,\mbox{rad}$.

\section{GRB Afterglows \& WSRT}
Observing the afterglow spectrum from X-ray to radio wavelengths enables us to construct the complete spectral energy distribution and follow its time evolution. The radio regime is especially useful as the temporal evolution of the afterglow at radio wavelengths is delayed in comparison with optical and longer wavelengths. This gives radio observations the unique possibility to study events that otherwise would escape attention, for example the passage of a reverse shock or that of the turn-over peak in the afterglow spectrum. Particularly, radio observations allow us to determine the self absorption frequency and the frequency that corresponds to the minimal energy in the electron energy distribution. Finally, the radio afterglow is visible much longer than optical and X-ray afterglows, so we can study the physics of the fireball in the non-relativistic phase.

The Westerbork Synthesis Radio Telescope (WSRT) is unique in the fact that there are very few telescopes which can detect the afterglow in the centimeter region, as these afterglows are extremely weak at these wavelengths. The closeness of the GRB\,030329 (with a redshift of $z = 0.1685$) caused a plethora of follow-up observations at all wavelengths. We observed this afterglow with the WSRT at 1.4, 2.3 and 4.8 GHz (21, 13 and 6 cm), starting one day after the burst, and we are still observing it. The observations show the peak frequency moving to longer wavelengths, eventually falling below the observation frequency and self absorption frequency. At the same time, one can find evidence for the jet nature of the afterglow, as the radio light curves change shape in a way that cannot be accounted for by an isotropic outflow. From early as well as late-time radio observations we can distinguish between models in which the circumburst medium has a constant density or a density gradient.

\section{Modeling GRB\,030329}
\begin{figure}
\includegraphics[width=13cm]{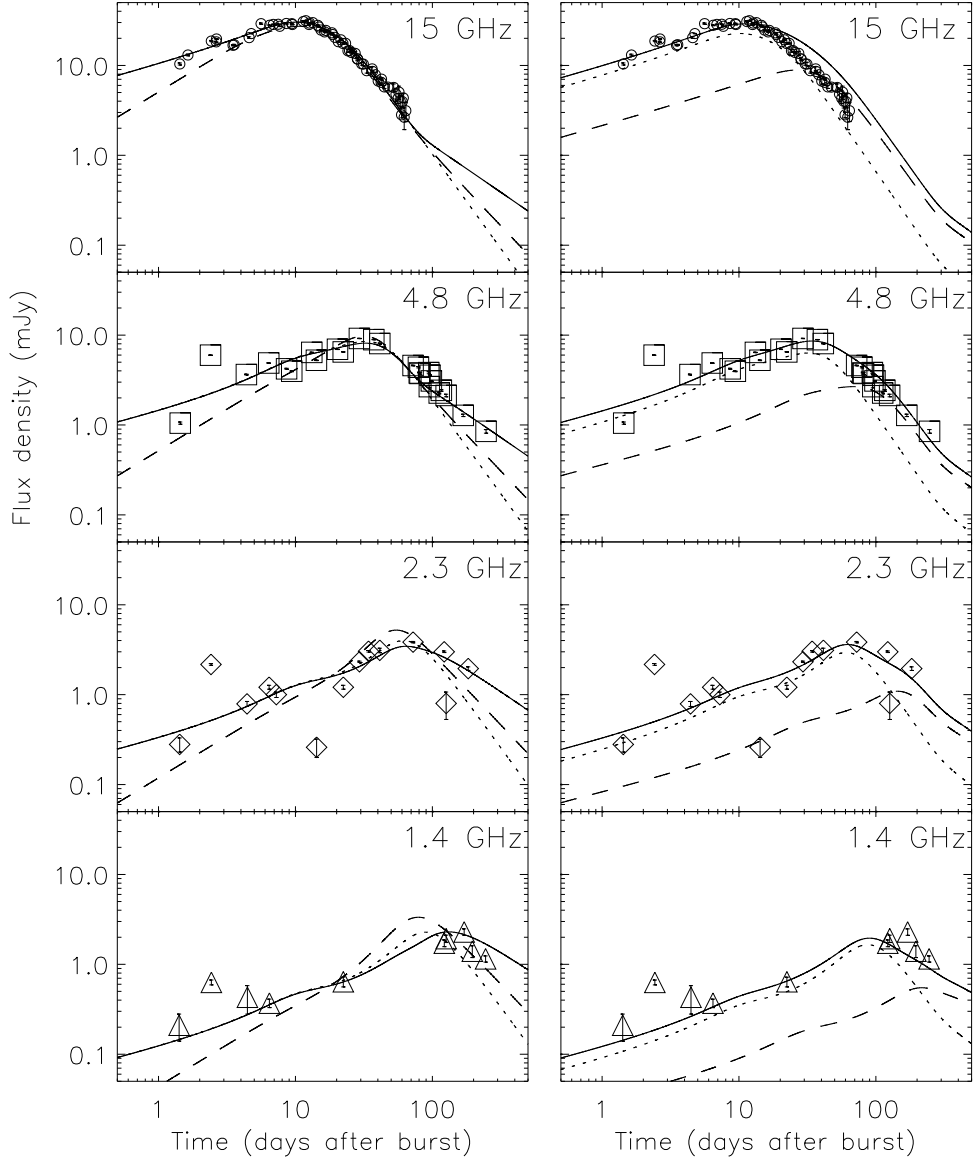}
\caption{The 1.4, 2.3 and 4.8 GHz WSRT light curves, with VLA/ATCA/RT 15 GHz observations \cite{ref:3} for comparison. {\bf Left:} The lines represent models with a peak frequency around 40 GHz and a self absorption frequency around 15 GHz at 10 days, when the jet starts to become visible (the so-called jet break time). The peak frequency falls below the self absorption frequency at 17 days. From then on, the maximum of a light curve at a given wavelength marks the passing of the self absorption frequency. The solid line corresponds to a model in which the fireball expands into a homogeneous medium and the non-relativistic phase of the fireball evolution starts after 80 days; the dotted line corresponds to the same model but without a non-relativistic phase; the dashed line corresponds to a model with a non-relativistic phase after 80 days and expansion of the fireball into a massive stellar wind. {\bf Right:} An extra wider jet with a lower Lorentz factor (right figure). The first component (dotted line, with a jet break time of 10 days) is responsible for the light curves until 50 days, while the second component (dashed line, with a jet break time of 30 days) accounts for the later peak in the light curves. The combined light curve is shown as the solid line.\label{fig:1}}
\end{figure}

The 1.4, 2.3 and 4.8 GHz WSRT light curves of the GRB\,030329 afterglow \cite{ref:2} are shown in Fig.\ref{fig:1}, together with VLA/ATCA/RT 15 GHz observations \cite{ref:3}. For this dataset we did not apply a $\chi^2$ fit, but tried to obtain a best fit by eye. This ignores the scatter in the data, which especially at early times is due to scintillation rather than measurement errors, and it will also put some more emphasis on the 1.4 GHz light curve. From observations at higher radio frequencies it was deduced that a jet with a jet-break time of $\approx 10$ days can explain the light curves \cite{ref:3}. Late time WSRT observations show an additional flux at the lowest frequencies after 50 days compared with the results obtained at higher radio frequencies. There are two possible explanations for this discrepancy:
\begin{itemize} 
\item A transition to non-relativistic expansion of the blastwave at $\sim 80$ days. The presence of a non-relativistic transition at $\sim 50$ days was predicted based upon X-ray data \cite{ref:4}, and observed with VLA radio data \cite{ref:5}. The non-relativistic expansion phase was studied in the cases of GRB\,970508 and GRB\,980703 \cite{ref:6}, and from that one can deduce the energy of the GRB independently of jet collimation. For GRB\,030329 more observations at later times are needed to do this. 
\item An extra wider jet with a lower Lorentz factor. The first component is responsible for the light curves until 50 days, while the second component accounts for the later peak in the light curves. Note that there is a smaller third component with a jet-break time of $0.55$ days that gives the early-time emission observed at short wavelengths (from the optical to gamma-ray bands) \cite{ref:7}. 
\end{itemize}

The two-component jet model is not satisfactory: it gives good fits at WSRT frequencies, but does not give a good fit to the data at radio frequencies above 4.8 GHz at late times. Our model in which a transition to a non-relativistic phase occurs after 80 days gives a better broadband fit except for the data at 1.4 GHz. Continuation of observations at late times at low radio frequencies can diagnose the cause of these discrepancies more closely.

In order to do broadband modeling of the GRB\,030329 afterglow one needs to extract the contribution from the supernova associated with this GRB at optical frequencies \cite{ref:8}\cite{ref:9}. Since the exact contribution from the supernova to the total flux in the optical is not known and because the supernova does not contribute significantly at low radio frequencies, the radio observations play a crucial role in doing proper broadband modeling and thus in determining the physical parameters very accurately.

\acknowledgments
We thank Richard Strom, Lex Kaper and Chryssa Kouveliotou for useful discussions. The Westerbork Synthesis Radio Telescope is operated by the ASTRON (Netherlands Foundation for Research in Astronomy) with support from the Netherlands Foundation for Scientific Research (NWO). We greatly appreciate the support from the WSRT staff in their help with scheduling the observations of GRB\,030329 as efficiently as possible.

\end{document}